**Elsevier Editorial System(tm) for Atmospheric Environment Manuscript Draft**

**Manuscript Number: ATMENV-D-09-01440R3**

Corresponding Author: Dr. Elena Chamizo,
Corresponding Author's Institution: Centro Nacional de Aceleradores (CNA), University of Seville
First Author: Elena Chamizo Order of Authors: Elena Chamizo; Manuel García-León; Santiago Miguel Enamorado; María del Carmen Jiménez-Ramos; Lukas Wacker

Abstract: Since the last nuclear atmospheric test carried out by the People Republic of China in 1980 and since the Chernobyl accident in 1986, the plutonium hasn't been directly released into the atmosphere. However, nowadays, it is still present in the troposphere. This is due to plutonium bearing soil particles physical resuspension processes. In this work, we study for the first time the temporal variation of plutonium isotopes, 239Pu and 240Pu, baseline concentrations on a monthly basis in surface air from Seville (Spain), and their correlation with some tracers of mineral dust, during 2001 and 2002. The Pu analyses were performed by low-energy Accelerator Mass Spectrometry (AMS). The 239+240Pu activity levels achieved maximums during the summer period, characterized by the absence of rains, and minimums during the rainy seasons, laying in the range 1-20 nBq*m-3. The 240Pu/239Pu two-year average atomic ratio was 0.18 ± 0.03, in agreement with the fallout plutonium. A good correlation with Pu and Al and Ti levels is observed. They are crustal components usually used as tracers of African dust over European countries. The hypothesis of the influence of the Saharan dust intrusions is supported as well through the study of Total Ozone Mass Spectrometer (TOMS) daily images.

**Abstract**

Since the last nuclear atmospheric test carried out by the People Republic of China in 1980 and since the Chernobyl accident in 1986, the plutonium hasn't been directly released into the atmosphere. However, nowadays, it is still present in the troposphere. This is due to plutonium-bearing soil particles physical resuspension processes. In this work, we study for the first time the temporal variation of plutonium isotopes, $^{239}$Pu and $^{240}$Pu, baseline concentrations on a monthly basis in surface air from Seville (Spain), and their correlation with some tracers of mineral dust, during 2001 and 2002. The Pu analyses were performed by low-energy Accelerator Mass Spectrometry (AMS). The $^{239}$Pu plus $^{240}$Pu ($^{239+240}$Pu) activity levels achieved maximums during the summer period, characterized by the absence of rains, and minimums during the rainy seasons, laying in the range 1-20 nBq·m$^{-3}$. The $^{240}$Pu/$^{239}$Pu two-year average atomic ratio was 0.18 ± 0.03, in agreement with the fallout plutonium. A good correlation with Pu and Al and Ti levels is observed. They are crustal components usually used as tracers of African dust over European countries. The hypothesis of the influence of the Saharan dust intrusions is supported as well through the study of Total Ozone Mass Spectrometer (TOMS) daily images.

## 1. Introduction.

The plutonium was synthesized for the first time in 1940. Since then, it has been massively introduced into the environment, due to military and civil uses of the nuclear energy. Well-known plutonium sources are the nuclear atmospheric tests carried out between 1945-1980; nuclear accidents such as the ones that took place in Chernobyl, Ukraine, in 1986, in Palomares, Spain, in 1966, or in Thule, Greenland, in 1968; and the controlled disposal (Sellafield, UK; La Hague, France) or accidental release (Mayak, Russia) of nuclear reactor wastes (UNSCEAR, 2000).

To date, the behaviour of Pu in the marine and terrestrial environments is well understood. However, the published information on its former and current distribution in the atmosphere is very scarce. During the years of the nuclear atmospheric tests, the distribution of atmospheric plutonium was governed by the air mass exchange between stratosphere and troposphere. The debris from the thermonuclear tests, injected directly into the stratosphere, would reach the troposphere mainly during the spring season after a half residence time of 11 months, being finally precipitated according to the meteorological conditions. This trend stopped by about 1985, once the contribution of the so-called "global fallout plutonium" became negligible, coming to light specific local effects related to nuclear accidents and Pu-bearing particles physical resuspension processes (wind erosion or anthropogenic activities) (Pan and Stevenson, 1995).

In the last three decades, the characterisation of the two major Pu isotopes, $^{239}$Pu ($T_{1/2}$= 24 110 y) and $^{240}$Pu ($T_{1/2}$=6 524 y), both alpha-emitters, in the air of populated areas affected by local plutonium sources, has been of major concern due to health reasons, as inhalation is considered one of the most important ways of incorporation of plutonium into the body (Nicholson, 1988). For instance, the case of the small town of Palomares, Spain, where the nuclear fuel of two thermonuclear bombs was accidentally dispersed in 1966, is tackled in (Iranzo et al., 1987); some results on the former US nuclear tests site in Nevada are reported in (Harley, 1980); the air monitoring of the surroundings of a nuclear waste isolation plant in New Mexico is assessed in (Arimoto et al., 2005); and the impact of the Chernobyl accident in the air of Prague, Czechoslovakia, in 1986, is studied in (Hölgye, 2008). However, very little is known about the current environmental baseline of plutonium in the air of areas free of local plutonium sources.

Some works point out to $^{239+240}Pu$ activities of about 1 $nBq·m^{-3}$ ($10^{-17}$ $g·m^{-3}$), several orders of magnitude lower than the reported ones for contaminated areas (Table 1), and to an influence of the presence of long-range transport of mineral dust (Choi et al., 2006). Analytical techniques offering very low detection limits are therefore necessary to perform general studies on a good temporal basis. Mass Spectrometry techniques such as ICP-MS and AMS are good candidates for this purpose. They offer as well the possibility of obtaining information on the $^{240}Pu/^{239}Pu$ atomic ratio, which is closely related to the contamination source. Recently, compact AMS systems working with terminal voltages lower than 1 MV have demonstrated to be competitive tools in this field (Wacker et al., 2005; Chamizo et al., 2008a).

In this work, we study for the first time the presence of $^{239,240}Pu$ in air filter samples from the city of Seville (Spain), taking advantage of the low detection limits offered by two multi-elemental compact Accelerator Mass Spectrometry (AMS) systems: the 600 kV facility at the Eidgenössische Technische Hochschule Zürich (ETHZ, Zürich, Switzerland), and the 1 MV system at the Centro Nacional de Aceleradores (CNA, Seville, Spain). The work was motivated by different reasons: to provide new data on the $^{239,240}Pu$ baseline levels in ground-level air in an urban area, to know the atmospheric Pu cycle in an area seasonally affected by Saharan dust intrusions, following the work of (Choi et al., 2006), and to give new data on the filters, routinely analysed for naturally-occurring and anthropogenic gamma-emitters (Jiménez-Ramos et al., 2006). The Pu data are complemented with the analysis of crustal components in the filters by ICP-MS. The sampling methodology, the Pu-AMS measurement procedure, and the obtained results are discussed in the following sections.

## 2.- Sampling conditions.

Since 1999, air filter samples are collected on the roof of the building of the Faculty of Physics of the University of Seville (37º21’33’’N; 5º59’12’’W), at 30 m above ground, within an air monitoring of radioactivity programme of the Spanish Nuclear Safety Council (CSN, Consejo de Seguridad Nuclear) (Figure 1). Sampling is carried out with the use of a high-volume air sampler (Aerosol Sampling Station, ASS-500), with a mean flow rate of 950 $m^3 \cdot h$. The aerosols are collected weekly on a 40x40 $cm^2$ polypropilene filter. After the routine determination of some naturally occurring ($^{7}Be$, $^{214}Pb$, $^{214}Bi$) and artificially (fission products, $^{59}Fe$, $^{95}Nb$, $^{131}I$, $^{137}Cs$) gamma-emitter radionuclides, the filters are stored for other scientific purposes (Jiménez-Ramos et al., 2006).

For the Pu-AMS determinations, the samples corresponding to two consecutive years were selected. 2001 and 2002 were chosen for two reasons: the existence of $^{129}I$ data, obtained as well by AMS, for the first period (Santos et al., 2006); and the occasional detection of $^{137}Cs$ in the routine gamma analyses during some summer months of those years. As, in this first approach, we wanted to obtained monthly averages, 10% aliquots of the individual filters were cut, grouped with that temporal resolution, and processed that way. On average, about 30 000 $m^3$ of filtered air and 6 g of aerosols were devoted per analysis. For the ICP-MS determinations, 1% aliquots of the weekly filters were processed following the same criterion.

## 3.- Sample preparation.

### 3.1.- Pu-AMS measurements.

In general, AMS determinations involve the chemical isolation of the element to be studied from the sample and its adjustment to a specific physical-chemical medium. In our case, the plutonium, once chemically isolated, has to be dispersed as $Pu_2O_3$ in an iron-oxide and aluminium matrix for the final preparation of the AMS cathode. This necessary procedure is usually performed in different steps. Firstly, a spike is added, usually $^{242}$Pu. That way, the original $^{239,240}$Pu concentrations can be finally quantified by the so-called isotope-dilution method. Secondly, the samples are burned at 600ºC in a muffle furnace. The resulting ashes are then leached in a hot plate with inorganic acids, and the plutonium fraction is finally purified from the supernantant by ion-chromatography methods (TEVA® (Eichrom Industries, Inc.) or AG®1X8 (Biorad Laboratories, Inc.) resins). About 1 mg of iron is then added to the plutonium solution as a carrier. This solution is baked to dryness, and the resulting residue is oxidized at 800ºC, mixed with aluminium powder in a 1:1 mass ratio, and pressed in a 1.3 mm diameter aluminium cathode.

For the preparation of the air-filter samples, the published procedure in (Chamizo et al., 2008b) was adopted with some specific modifications. First of all, to avoid burning the polypropylene filter matrix, the samples were converted to ashes with a special temperature ramping program: 30ºC per hour till 300ºC -the degradation temperature of the polypropylene -, and 80ºC per hour till the final set value, 600ºC. Both temperatures were maintained for about 4 hours. On these ashes, the $^{242}$Pu spike was then added. That way, it could be properly homogenized. Any special modifications were introduced in the following steps: the leaching was performed with concentrated analytical-grade $HNO_3$ and $H_2O_2$, and the isolation of the plutonium fraction was achieved with 2 ml prepacked TEVA® resins, following the adjustment of the plutonium to Pu(IV) with a two-step redox process (Chamizo et al., 2008b).

The Pu-AMS measuring procedures at the ETH (Zürich) and at the CNA (Seville) facilities have been detailed in (Wacker et al., 2005) and (Chamizo et al., 2008a), respectively. Briefly, the plutonium is extracted from the cathode as $PuO^-$ in a Cs sputter ion source, stripped to $Pu^{3+}$ in Ar gas at the accelerator terminal with about 11% yield, and finally counted from the total energy signal provided by a gas ionisation chamber with a 30 nm thickness silicon nitride window. Pulsing times of a few seconds are dedicated to every isotope ($^{239,240,242}Pu$), adding up to about 30 minutes of analysis per sample. The typical instrumental error is about 2%. Figures of merit of this technique are the backgrounds -of about 5 µBq for both $^{239,240}Pu$ based on procedural blanks-, and the $^{239}Pu/^{238}U$ mass suppression factor – higher than $10^{-7}$. The drawback is the poor ionisation efficiency for the Pu achieved in the sputtering process (Child et al., 2009). A summary of the whole measuring procedure in the case of the CNA system is given in Figure 2.

**3.2.- ICP-MS determinations.**

The study of the major elements in the filters was carried out on the ICP-MS facility (Thermo Elemental X-7) at the building of the School of Agricultural Engineering (EUITA, Escuela Universitaria de Ingeniería Agrícola ) of the University of Seville. The necessary dissolution of the sample was performed by microwave digestion. About 8 ml of Suprapur $HNO_3$ at 65% and 4 ml of MQ® water were used per 300 mg filter and reactor. An Anton-Paar MW oven, model MX-3000, provided with 16 reactors, was used. The leaching was performed following a temperature program: firstly, the temperature was raised to 120ºC in 30 min and maintained for 20 min; secondly, it was increased to 180ºC and kept for 40 min more. This process was repeated as needed until the solution was clear. Finally, the solution was heated to dryness and dissolved in 20 ml of 2% $HNO_3$ for the ICP-MS determinations.

**4.- Results.**

The obtained $^{239,240}Pu$ results ($^{239+240}Pu$ activity concentrations, in $nBq\cdot m^{-3}$, and $^{240}Pu/^{239}Pu$ atomic ratios, in %) for the two studied years with monthly resolution are depicted in Figure 3 and shown in Table 2. The reproducibility of these analyses was demonstrated by analysing second aliquots of three samples, whose results are displayed as well in the Table 2.

As it can be seen, the plutonium in the air of Seville follows a seasonal cycle, with maximums during the summer and minimums during the rainy months. Specifically, the maximum $^{239+240}Pu$ activity concentrations were achieved in August 2001 (18.82 ± 0.83 $nBq\cdot m^{-3}$) and in June 2002 (14.66 ± 0.75 $nBq\cdot m^{-3}$), and the minimum ones in January 2001 (1.95 ± 0.17 $nBq\cdot m^{-3}$) and in November 2002 (1.72 ± 0.23 $nBq\cdot m^{-3}$). They are in agreement with the reported atmospheric levels in Madrid, Spain, in 2005 (Gascó et al., 2007) (Table 3). Most of the $^{240}Pu/^{239}Pu$ atomic ratios are in agreement with the reported ones for the global fallout plutonium in the Northern Hemisphere, whose integrated average is (18 ± 1.4)% (Kelley et al., 1999) (Table 3). However, there are some effects that will be discussed in the next section. On the other hand, it is interesting to point out that a $^{137}Cs$ signal was clearly identified during the first week of August 2001, of 1.40 ± 0.25 $\mu Bq\cdot m^{-3}$. In 2002, the $^{137}Cs$ signal couldn’t be quantified, although it exceeded the limit of detection during June and July.

## 5.- Discussion.

The obtained plutonium seasonal cycle in the air of Seville is in agreement with the soil resuspension hypothesis. Indeed, Seville hasn´t been affected by any local plutonium source, so the presence of this anthropogenic element in its atmospheric samples can only be explained considering the injection of Pu-bearing soil particles into the troposphere by dust storms. It explains the existence of a positive correlation between the $^{239+240}$Pu activity concentrations and the total mass of aerosols in the filters (Figure 4), which improves when the concentrations of some typical crustal elements, as determined by ICP-MS, are considered in the fits: among them, the ones showing the best correlation with Pu are, by decreasing order, Be ($r^2$=0.7680), Co ($r^2$=0.7558), Al ($r^2$=0.7557), Cs ($r^2$=0.6894) and Ti ($r^2$=0.6402). As an example, the corresponding linear fits for Be, Al and Ti are displayed in Figures 5, 6 and 7, respectively. Therefore, that way we confirm the hypothesis of the soil resuspension as a source of present atmospheric plutonium. The open questions are now the origin of the mineral dust and the reason why the $^{239+240}$Pu activity maximums are achieved in August 2001 and in June 2002.

In order to distinguish between local-soil resuspension processes and the influence of long-range transport episodes of mineral dust, one of the most wide-spread methods is the visual analysis of the Total Ozone Mass Spectrometer (TOMS) daily images available online through the National Aeronautics and Space Administration (NASA) webpage. They inform about the relative turbidity of the atmosphere through a parameter called Aerosol Index. This parameter is determined by measuring spectral intensities on 340 and 380 nm lines in backscattered radiation and comparing experimental data with calculated ones deduces from a reflectivity model (Borbély-Kiss et al., 2004). This information, static, can be complemented with the synoptic data provided by the Dust Regional Atmospheric Model (DREAM), a tool offered by the Barcelona Supercomputing Center, for instance. As an example, in Figure 8 we show the TOMS and DREAM images for the 17$^{th}$ of June 2002, a day where the intrusion into the South-west of Spain of an air mass from the North of Africa can be clearly identified.

Analysing the daily TOMS images for the two studied years, Saharan Dust Intrusions (SDI) were identified during 9 months in 2001 and 6 months in 2002. They were accompanied by wind blowing south-west (Instituto Nacional de Meteorología, INM), in agreement with the information obtained from the DREAM model. In Figure 9, the average value of the monthly Aerosol Index, the rainfall as reported by the INM, and the Pu massic activities in surface air, in $mBq \cdot g^{-1}$, are plotted for the 24 studied months. As it can be observed, the seasonal trend of $^{239,\,240}Pu$ in the air of Seville seems to be modulated by SDI, especially in 2002. However, additional experimental evidences support this hypothesis for the two studied periods. Firstly, the same SDI had been already identified by other authors working on the physical-chemical speciation of aerosols in different parts of Spain, with also maximum crustal dust content in August 2001 and June 2002, in agreement with our results (Querol et al., 2005; Salvador et al., 2007). Secondly, the maxima $^{239+240}Pu$ massic activities, of 0.160 $mBq \cdot g^{-1}$, the same for August 2001 and June 2002, could be explained considering the $^{239+240}Pu$ levels in desert areas from Northern Africa, as they are in agreement with the published levels for soils from Southern Algeria, affected by the general fallout, ranging from 0.1 to 0.6 $mBq \cdot g^{-1}$ (Baggoura et al., 1998). Another experimental evidence may be the $^{137}Cs/^{239+240}Pu$ activity ratio measured in August 2001. This is about 80 (decay corrected to 2010), in agreement with the reported levels for soils from Algeria, influenced by the Chernobyl accident, from 30 to 100 (Baggoura et al., 1998), but significantly higher than the one for the current general fallout, from 15 to 30 (PNUE/IAEA, 1992). Finally, the $^{240}Pu/^{239}Pu$ monthly atomic ratios, which are lower than the expected one for the integrated fallout in the Northern Hemisphere during the months where no SDI were identified (Figure 10). This decreasing trend could be explained considering the presence of local soil particles, with a characteristic $^{240}Pu/^{239}Pu$ atomic ratio lower than the previous one, of (14.7 ± 0.8)%, as determined in this work from the study of a set of local samples. A similar ratio had been already reported for Sacavem, Portugal, of (16.55 ± 0.09)%, in (Kelley et al., 1999). However, during three months the obtained ratios seems to follow a different behaviour (a, b and c in Figure 10), possibly showing specific local effects. The higher values could be explained considering the presence of soil particles from Northern Algeria, which was affected by the Chernobyl accident (Baggoura et al., 1998). However, due to the poor statistics and to the scarce data, additional analyses are necessary to explain them satisfactorily.

## 6.- Conclusions.

There are experimental evidences showing that the $^{239,240}Pu$ activity levels in the air of Seville, Spain, are modulated by Saharan dust intrusions. A similar effect had been already identified in Daejeon, South Korea, considering the entrainment of soil particles from the Northern China deserts (Choi et al., 2006).

## 7.- Acknowledgements.

This work has been financed through the project FIS2008-01149/FIS of the Spanish Ministry of Science and Innovation. The authors are indebted to the INM for providing the meteorological data, and to Applied Nuclear Physics group of the University of Seville for providing the air filter samples.

## 8.- Bibliography.

## 8.- Figure Captions.

Figure 1.- Location of the building of the Faculty of Physics of the University of Seville, where the air filter samples were collected, and of the meteorological stations of the "Instituto Nacional de Meteorología" (INM).

Figure 2.- On the left side, summary of the main steps of the sample preparation procedure. On the right side, sketch of the 1 MV AMS system at the CNA together with the Pu measurement details. The picture below the ion source corresponds to the surface of the cathode where the sample of interest is dispersed. As it can be seen in the final spectra, the $Pu^{3+}$ ions can be totally discriminated against the molecular fragments with the same M/q ratio, which pass the high-energy cinematic filters. For further details, see (Chamizo et al., 2008a).

Figure 3.- Monthly evolution during 2001 and 2002 of the $^{239+240}Pu$ activity concentrations (▲, left axis, in $nBq/m^3$) and the $^{240}Pu/^{239}Pu$ atomic ratios (♦, right axis, in %) in the air of Seville. The data points are shown in Table 2.

Figure 4.- $^{239+240}Pu$ activity concentration versus the total mass of the aerosols (TSP, Total Suspended Particles) in the air of Seville for the two studied years.

Figure 5.- $^{239+240}Pu$ activity concentration versus the beryllium concentration in the air of Seville for the two studied years.

Figure 6.- $^{239+240}Pu$ activity concentration versus the aluminium concentration, as determined by ICP-MS, in the air of Seville for the two studied years.

Figure 7.- $^{239+240}Pu$ activity concentration versus the titanium concentration, as determined by ICP-MS, in the air of Seville for the two studied years.

Figure 8.- Images available online through the NASA web page (a, TOMS images, http://toms-gsfc.nasa.gov/aerosols/) and through the Barcelona Supercomputing Center (b, DREAM model, http://www.bsc.es/projects/earthscience/DREAM) showing the dust content of the atmosphere worldwide over northern Africa and southern Europe, respectively, for the $17^{th}$ of June 2002.

Figure 9.- Monthly evolution of the $^{239+240}Pu$ massic activity in surface air (▲, left axis, in $nBq \cdot g^{-1}$), a parameter related to the average Aerosol Index in relative units (grey bars, right axis), and the rainfall (white bars, right axis), during 2001 and 2002 in Seville.

Figure 10.- $^{240}Pu/^{239}Pu$ atomic ratios versus the Al concentration in the monthly samples. The months where SDI were identified have been represented with bold symbols (■). The characteristic ratios for the general fallout in the Northern Hemisphere and the one determined in local soils from Seville, together with their corresponding errors, are displayed for comparison purposes (see discussion in the text).

## 9.- Table Captions.

Table 1.- Reported $^{239+240}Pu$ activity concentrations in the air of areas affected by different plutonium sources, obtained by alpha spectrometry. In every case, the sampling station was located between 2 and 10 m above ground.

Table 2.- $^{239+240}Pu$ activity concentrations and $^{240}Pu/^{239}Pu$ atomic ratios for the two studied years on a monthly basis. For the cases of July 2001, June 2002 and October 2002, the second lines correspond to second aliquots of the filters that were analysed to check the reproducibility of the results.

Table 3.- In the first and second columns: maximum and minimum $^{239+240}Pu$ activity concentrations ($A_{max}$ and $A_{min}$, respectively) for the air of Seville in 2001 and 2002, and Madrid in 2005 (Gascó et al., 2007). These last data were obtained by alpha-spectrometry. Third column: average $^{240}Pu/^{239}Pu$ atomic ratio for Seville for the two studied years. In the case of 2002, the data corresponding to December is not included in the calculations.

Figure 1

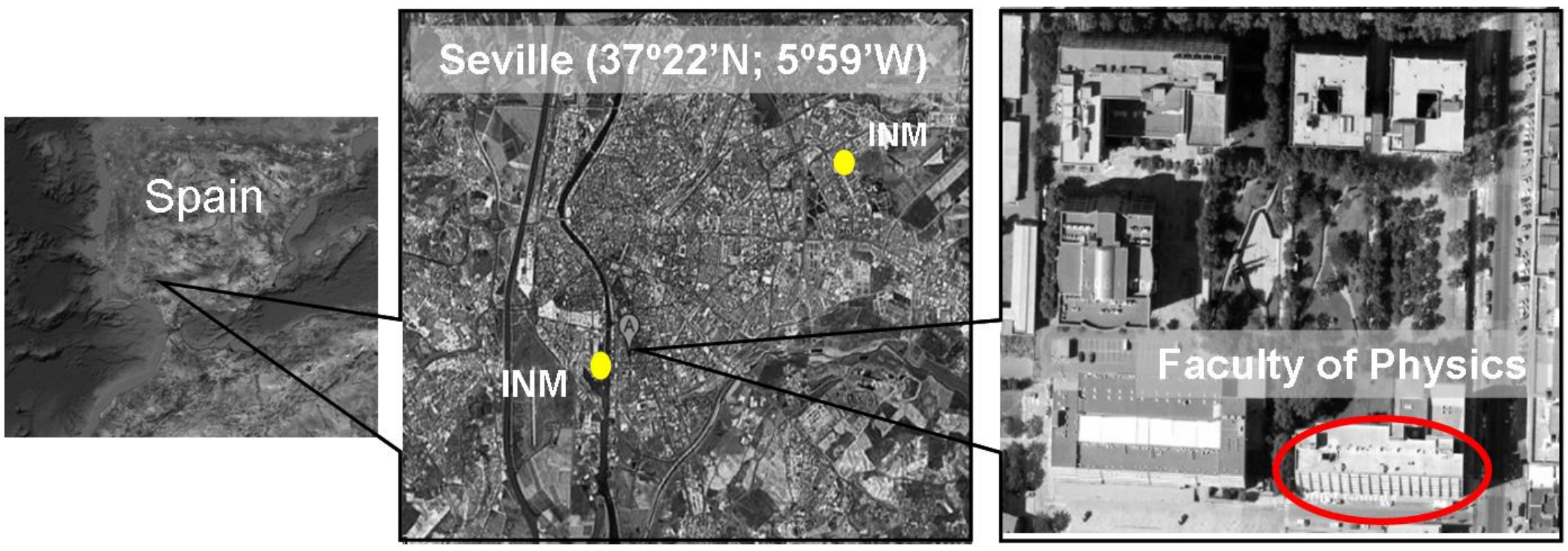

Figure 2

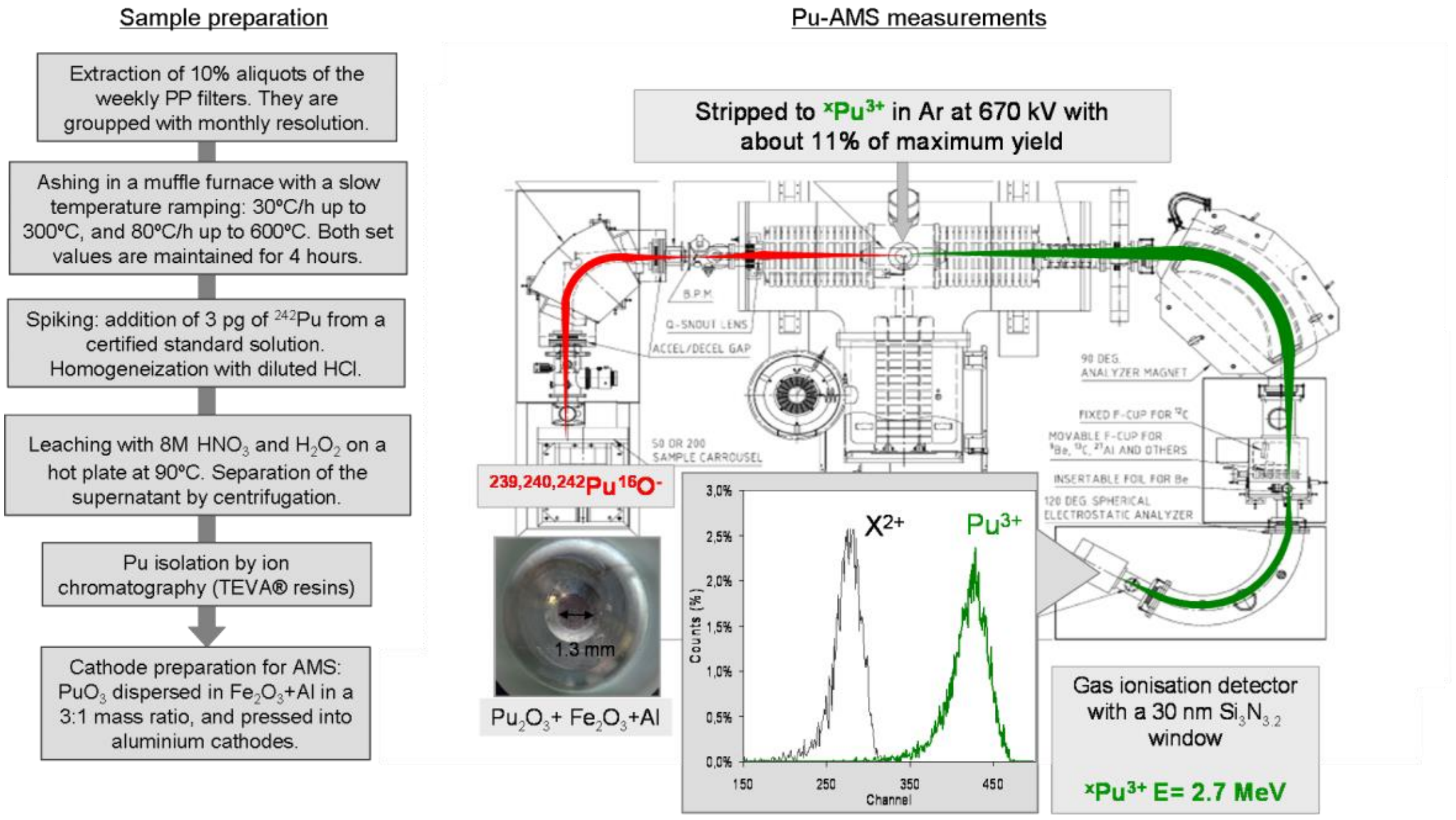

Figure 3

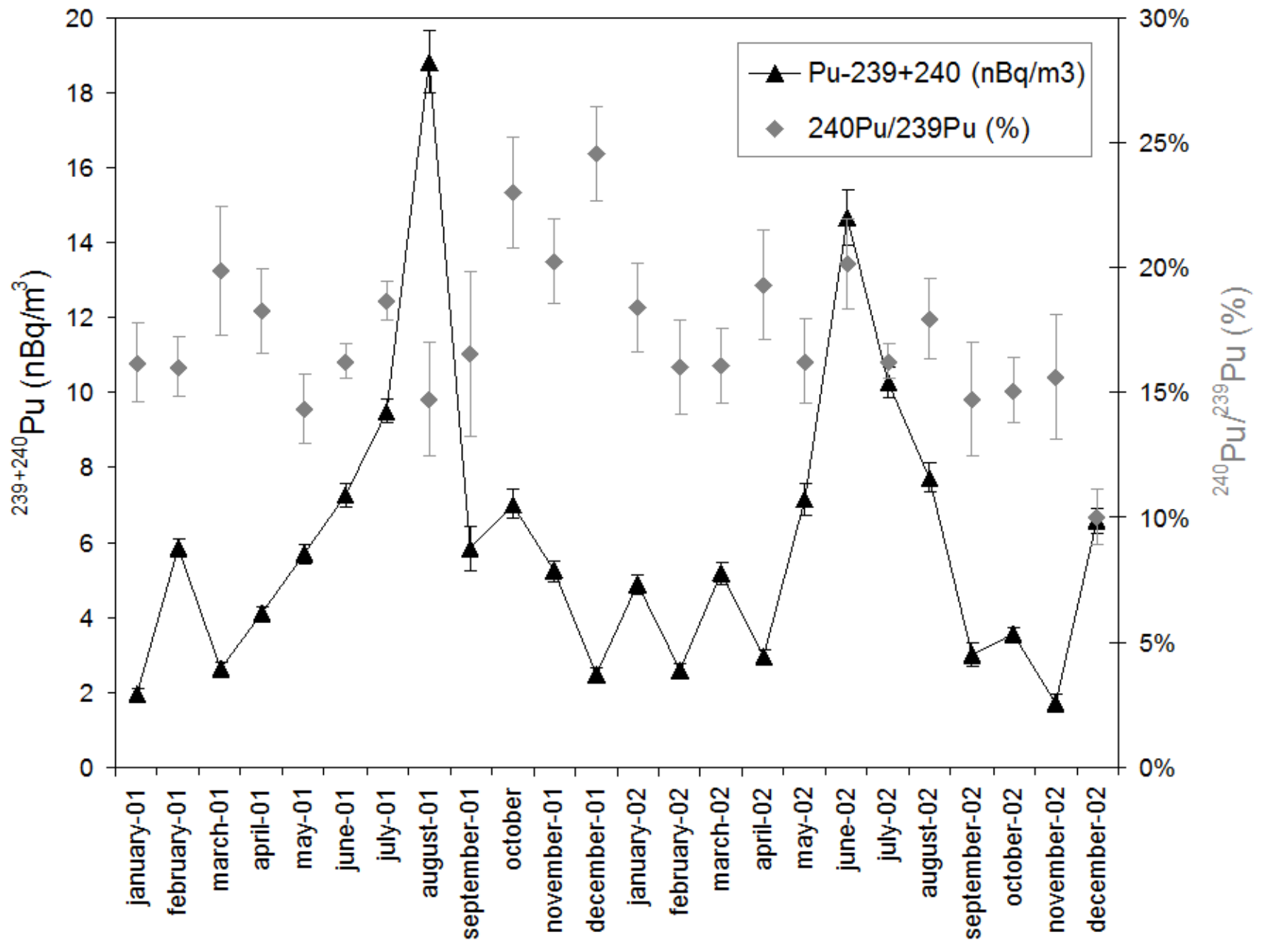

Figure 4

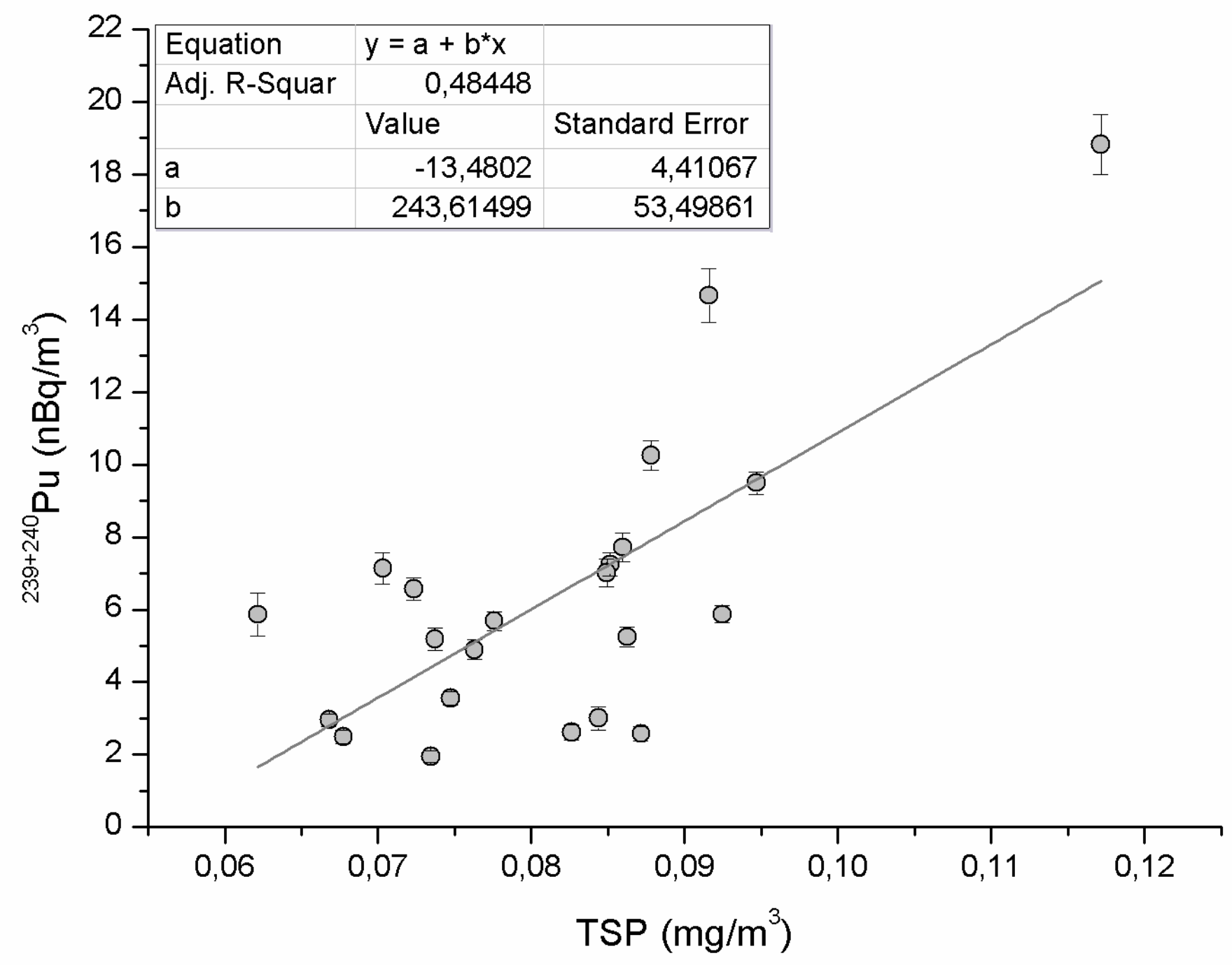

Figure 5

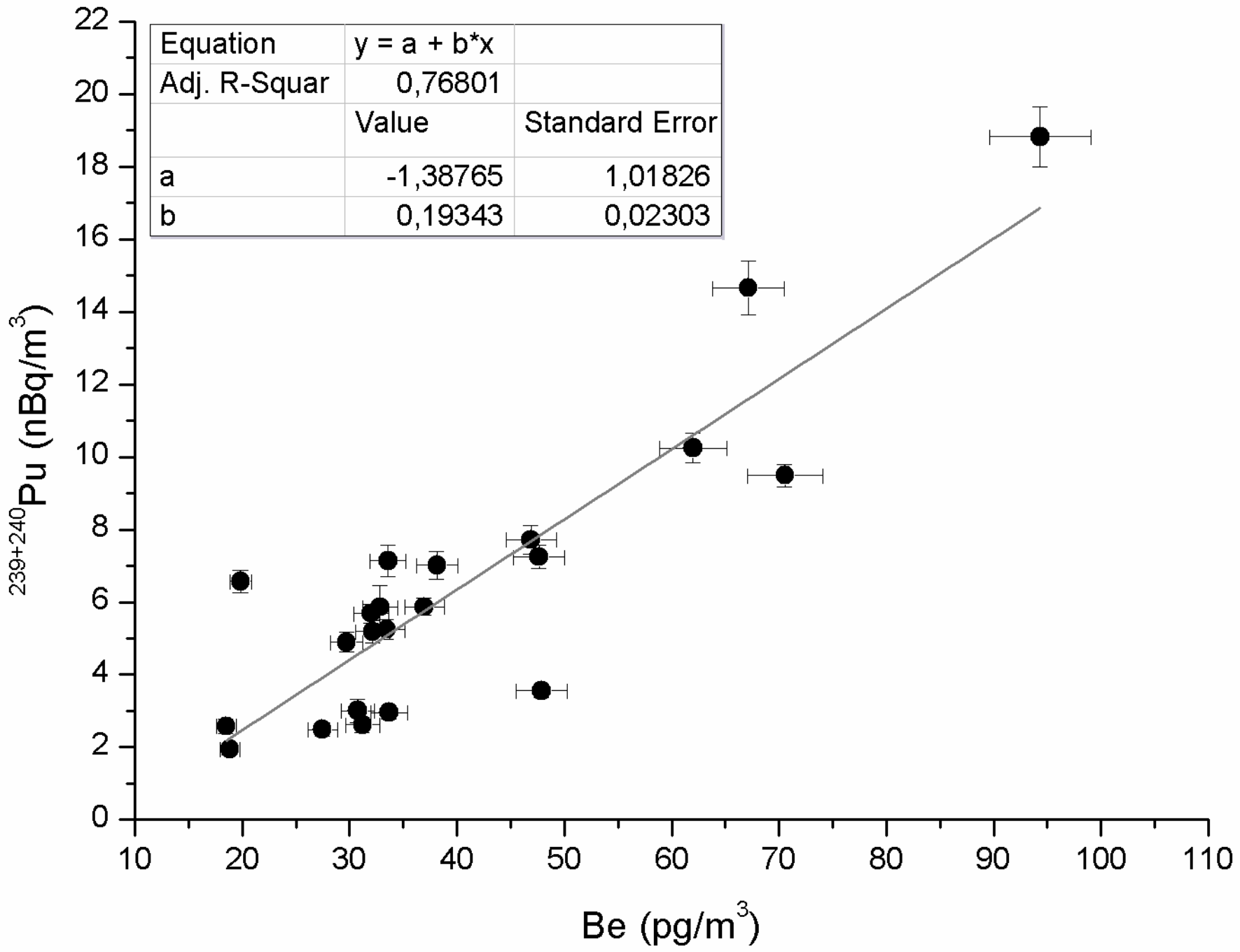

Figure 6

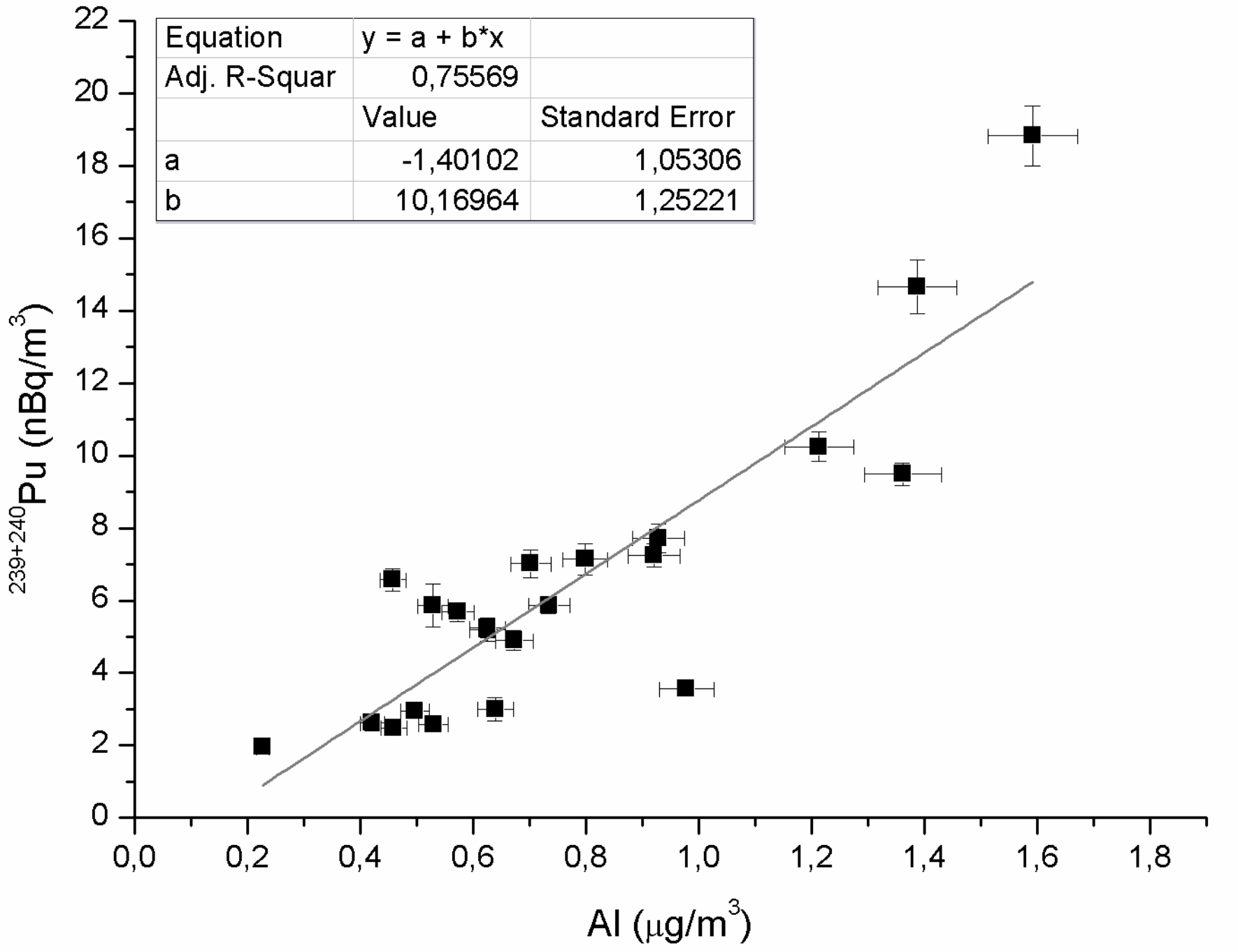

Figure 7

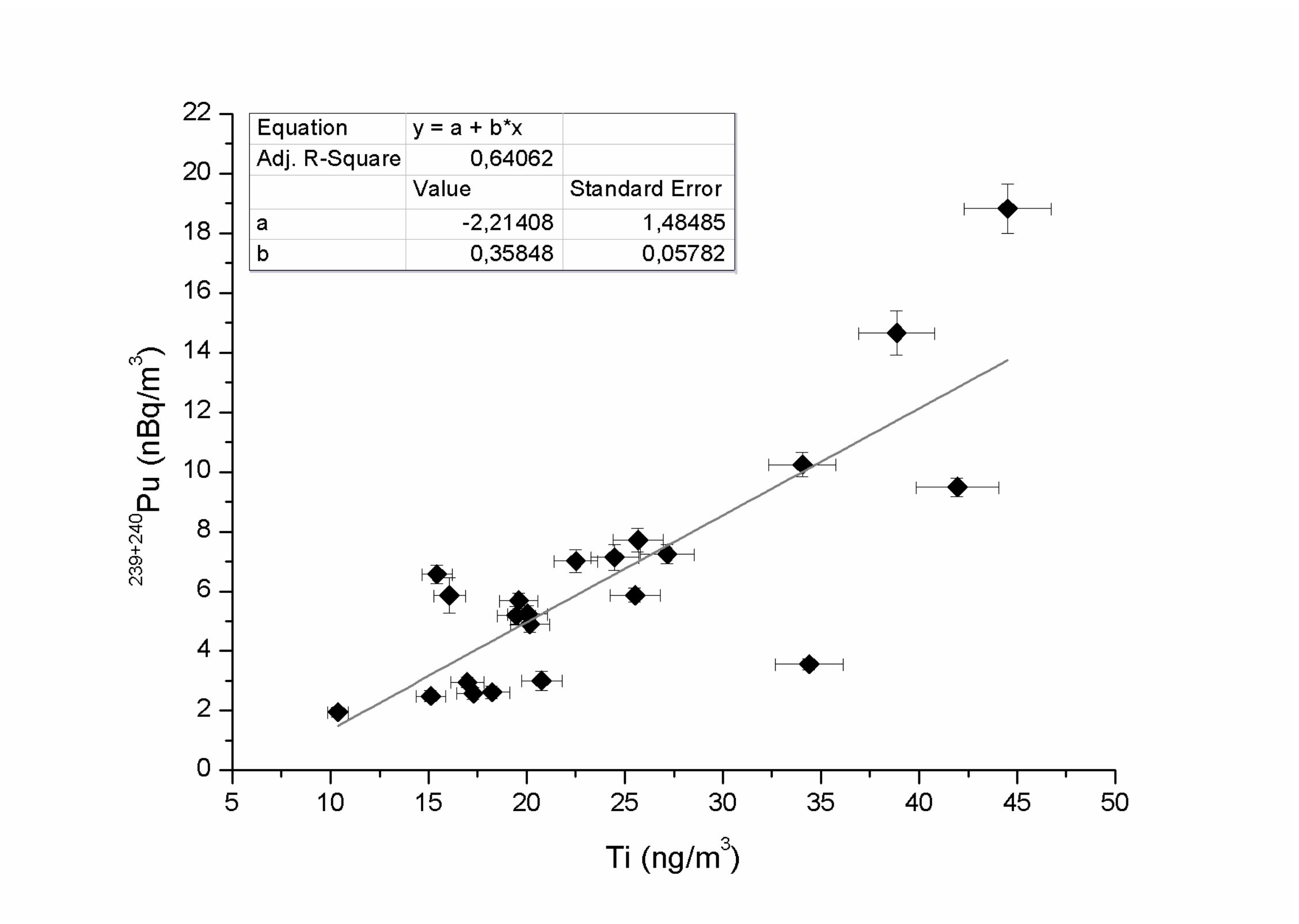

Figure 8

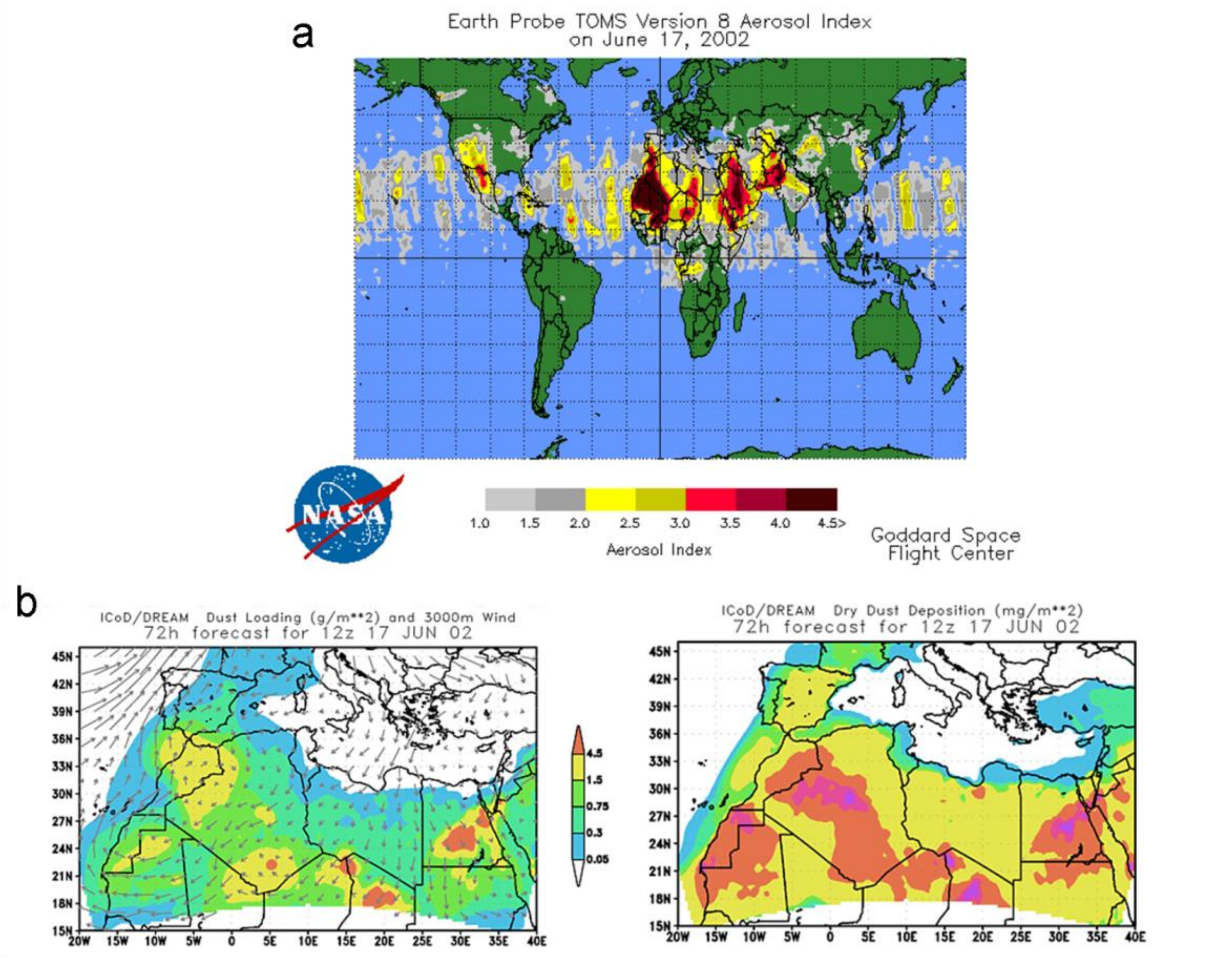

Figure 9

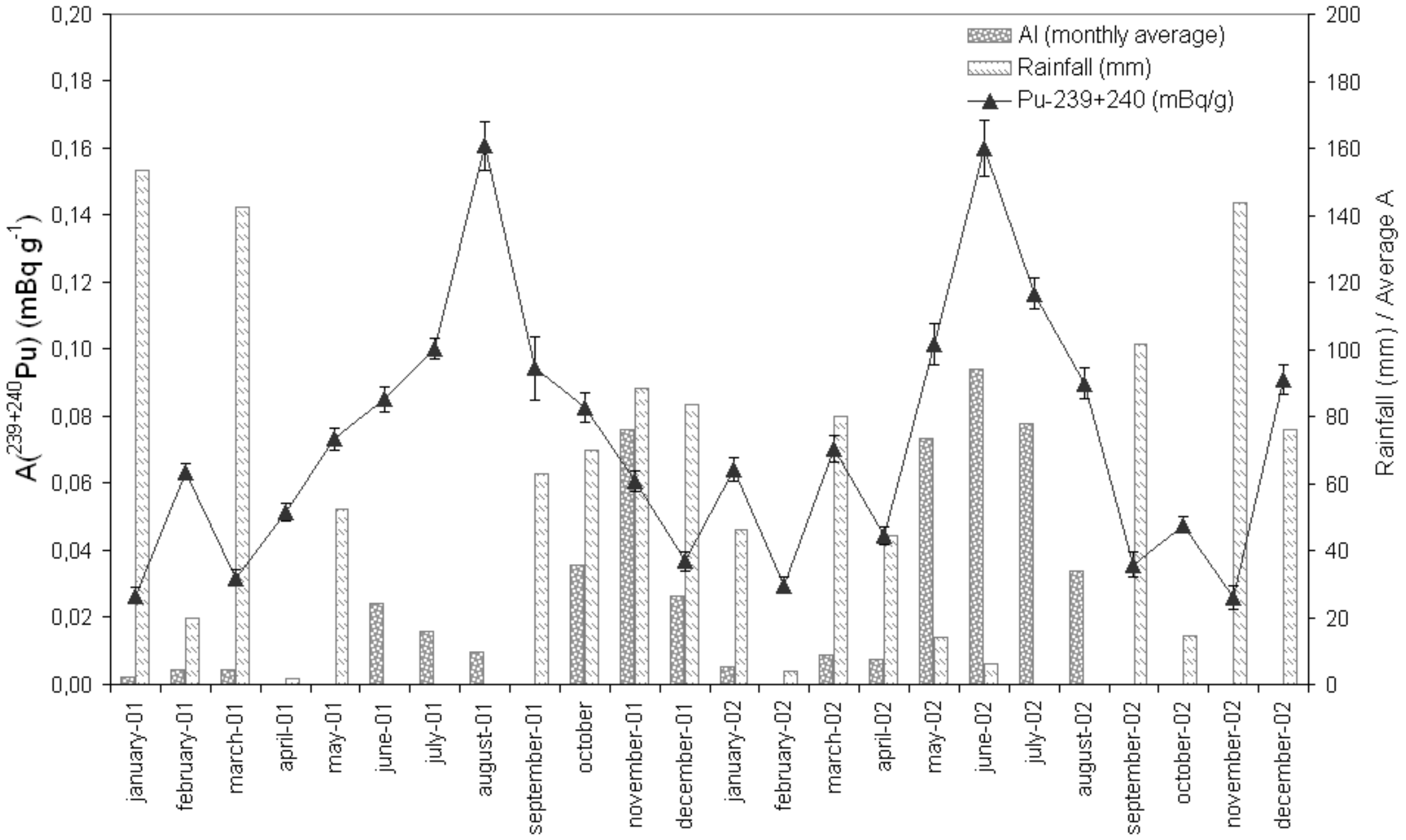

Table 1

| Sampling point | Plutonium source | A ($^{239+240}$Pu)·V$^{-1}$ | Reference |
|---|---|---|---|
| Nevada Test Site, Nevada, United States | Low-yield nuclear atmospheric tests between 1945 and 1961. | 4 mBq·m$^{-3}$ | (Harley, 1980) |
| Palomares, Almería, Spain | Nuclear accident in 1966. | 5-55 µBq·m$^{-3}$ | (Iranzo et al., 1987) |
| Prague, Czechoslovakia, | Chernobyl accident (Ukraine, 1986) | 10-30 µBq·m$^{-3}$ (1986) | (Hölgye, 2008) |
| Waste Isolation Pilot Plant, New Mexico, United States | Fallout | 15 nBq·m$^{-3}$ | (Arimoto et al., 2005) |
| Braunschweig, Germany | Fallout | 2 nBq·m$^{-3}$ | (Wershofen et al., 2001) |
| Chicago, United States | | 15 nBq·m$^{-3}$ | (Pan and Stevenson, 1996) |

Table 2

| | 2001 | | | 2002 | | |
|---|---|---|---|---|---|---|
| Month | Interval | $^{239+240}Pu$ ($nBq \cdot m^{-3}$) | $^{240}Pu/^{239}Pu$ (atomic ratio) | Interval | $^{239+240}Pu$ ($nBq \cdot m^{-3}$) | $^{240}Pu/^{239}Pu$ (atomic ratio) |
| January | 02-01/30-01 | 1.95 ± 0.17 | 0.162 ± 0.016 | 31-12/04-02 | 4.89 ± 0.27 | 0.184 ± 0.018 |
| February | 30-01/26-02 | 5.87 ± 0.23 | 0.160 ± 0.012 | 04-02/04-03 | 2.58 ± 0.20 | 0.160 ± 0.019 |
| March | 26-02/26-03 | 2.62 ± 0.20 | 0.199 ± 0.026 | 04-03/01-04 | 5.19 ± 0.30 | 0.161 ± 0.015 |
| April | 26-03/30-04 | 4.09 ± 0.20 | 0.183 ± 0.017 | 01-04/29-04 | 2.96 ± 0.17 | 0.193 ± 0.022 |
| May | 30-04/04-06 | 5.69 ± 0.26 | 0.143 ± 0.014 | 29-04/03-06 | 7.14 ± 0.43 | 0.162 ± 0.017 |
| June | 04-06/02-07 | 7.24 ± 0.32 | 0.162 ± 0.007 | 03-06/02-07 | 14.66 ± 0.75<br>14.48 ± 0.88 | 0.202 ± 0.018<br>0.171 ± 0.036 |
| July | 02-07/30-07 | 9.49 ± 0.31<br>10.28 ± 0.45 | 0.187 ± 0.008<br>0.185 ± 0.015 | 02-07/29-07 | 10.25 ± 0.41 | 0.162 ± 0.007 |
| August | 30-07/10-08 | 18.82 ± 0.83 | 0.147 ± 0.023 | 29-07/02-09 | 7.72 ± 0.39 | 0.179 ± 0.016 |
| September | 17-09/01-10 | 5.86 ± 0.59 | 0.165 ± 0.033 | 02-09/30-09 | 3.01 ± 0.32 | 0.147 ± 0.023 |
| October | 01-10/29-10 | 7.02 ± 0.38 | 0.230 ± 0.22 | 30-09/04-11 | 3.56 ± 0.19<br>2.94 ± 0.18 | 0.151 ± 0.013<br>0.192 ± 0.021 |
| November | 29-10/03-12 | 5.24 ± 0.27 | 0.202 ± 0.017 | 04-11/03-12 | 1.72 ± 0.23 | 0.156 ± 0.025 |
| December | 03-12/31-12 | 2.49 ± 0.19 | 0.245 ± 0.019 | 03-12/30-12 | 6.57 ± 0.31 | 0.100 ± 0.011 |

Table 3

| | $A_{max}$ ($^{239+240}Pu$) ($nBq \cdot m^{-3}$) | $A_{min}$ ($^{239+240}Pu$) ($nBq \cdot m^{-3}$) | $^{240}Pu/^{239}Pu$ (atomic ratio) | Reference |
|---|---|---|---|---|
| 2001 (Seville) | 18.82 ± 0.83 (August) | 1.95 ± 0.17 (January) | 0.18 (0.03) | This work |
| 2002 (Seville) | 14.66 ± 0.75 (June) | 1.72 ± 0.23 (November) | 0.17 (0.02) | This work |
| 2005 (Madrid) | 13 ± 3 (June and August) | 2 ± 1 (December) | ------- | (Gascó et al, 2007) |